\documentclass[pra,twocolumn,showpacs,aps,floatfix,amsmath,amssymb,amsfonts]{revtex4}
\pdfoutput=1
\usepackage{amsmath}
\usepackage{graphicx}
\usepackage{color}

\begin{document}
\title{Feshbach resonances in a nonseparable trap}
\author{Krzysztof Jachymski$^{1,2}$, Zbigniew Idziaszek$^{1}$ and Tommaso Calarco$^{2}$}
\affiliation{$^1$Faculty of Physics, University of Warsaw, Ho{\.z}a 69,
00-681 Warsaw, Poland,\\
$^2$Institut f\"{u}r Quanteninformationsverarbeitung, Universit\"{a}t Ulm, D-89069 Ulm, Germany}


\date{\today}

\begin{abstract}
We consider a pair of atoms in an arbitrary trapping potential in the presence of magnetically tunable Feshbach resonance. We find the energy levels and occupation of the bound molecular states taking into account possible coupling between center of mass and relative motion induced by the trap. As a specific example we discuss the case of different atomic species in harmonic potential, where each atom feels different trapping frequency.
\end{abstract}

\maketitle
\section{Introduction}
Degenerate quantum gases manipulated by electromagnetic fields provide an opportunity to perform quantum simulations of condensed matter models as well as quantum information processing~\cite{BlochRMP,LewensteinAdv,Cirac2005}. In experiments, it is possible not only to precisely control the trapping potential, but also to vary the interactions strength, described by the scattering length. This can be done using magnetically tunable Feshbach resonances, which are one of the most important phenomena in this field~\cite{JulienneRMP,Timmermans}. Feshbach resonances are a result of coupling between the free atom pair and a bound state in the closed channel. The position of this state can be controlled using an external magnetic field, causing a resonance in the scattering length when it crosses the threshold. A large scattering length greatly increases three-body losses in the system, which was the first sign of such resonances in experiment~\cite{Inouye}. Nowadays, Feshbach resonances are a crucial tool to produce ultracold molecules, which can be done e.g. by using time-dependent magnetic fields~\cite{Mies2000,Julienne2004,KohlerRMP}. Feshbach resonances allow also to observe atom-molecule coherence~\cite{Wieman}, BEC-BCS crossover~\cite{Bourdel,Grimm} and to simulate identical fermions with a pair of distinguishable atoms~\cite{Jochim}.

The simplest description of Feshbach resonances uses a single-channel model in which the atoms are assumed to interact via a pseudopotential with scattering length $a$ depending on the magnetic field~\cite{JulienneRMP}. This approach is limited to the so-called open-channel dominated resonances~\cite{JulienneRMP}. To increase the precision, especially for closed-channel dominated (also called narrow) resonances, more complex methods are needed. Multichannel calculations and experimental Feshbach spectroscopy have been performed for numerous cases (\cite{Julienne2000,Marzok2009,Schuster2012,Takekoshi2012,Weidemueller2012} and many more). Full coupled channel calculations can be simplified to effective two-channel models~\cite{Raoult2000,Kokkelmans2002,Schneider2009}.

In the presence of external harmonic traps, single-channel pseudopotential calculations can be performed analytically~\cite{Bush}, but their validity is limited to the case when the range of the potential is much smaller than the trap width and the scattering length $a$ is sufficiently small, so that $ka\ll 1$~\cite{Tiesinga2000}. Close to the resonance, when $a$ can be arbitrarily large, the description has to be extended by introducing an energy-dependent scattering length~\cite{Bolda2002,Bolda2003,Idziaszek2006}.

In free space and in a harmonic trap the center of mass and relative motion of a pair of atoms are decoupled, and the Feshbach resonance changes only the properties of the relative motion. However, for a variety of traps used in experiments, such as optical lattices or double wells, such separation is not possible anymore. This leads to novel phenomena, such as anharmonic confinement-induced resonances~\cite{Bolda2005,Kestner2010,Drummond}, and formation of states with nontrivial angular momentum correlations~\cite{Bertelsen2007}. Anharmonic terms influence the energy of the pair of atoms and the structure of bound states~\cite{Sengstock}, also in a waveguide~\cite{Melezhik2009} and in optical lattices~\cite{Dickerscheid2005,Buchler,Sala2012,DienerHo2006,Carr}.

In this work we consider a pair of atoms confined in an arbitrary external trapping potential in the vicinity of a~Feshbach resonance. We describe the resonance using a two-channel configuration interaction (CI) model. In our approach we treat the closed channel molecular state as a~pointlike particle. This approximation results in divergencies, which can however be renormalized, in close analogy to the free space problem~\cite{Kokkelmans2002}. A convenient way to do this is to introduce the renormalized resonance shift~\cite{Buchler,Carr}. We then analyze the Feshbach resonance in harmonic trap. In this case the renormalization procedure is particularly simple and can be implemented numerically without difficulties. This result makes it possible to perform efficient numerical calculations for the large class of traps where harmonic oscillator solutions can be used. As a simple example of a nonseparable problem, we describe association of a heteronuclear molecule. The separation of center of mass and relative motion does not occur, because atoms with different masses and polarizabilities feel different trapping frequencies~\cite{Bertelsen2007,Sengstock}. Because in experiments involving mixed species the atoms are always in external traps usually having different frequencies, taking the effects of nonseparability of the trap into account is crucial for the accuracy of calculations.

Our paper is organized as follows. In Section II we outline the basic physics of Feshbach resonances and the parameters which describe them. In Section III we generalise the theory to the case of arbitrary trapping potentials and obtain self-consistent equations for the energy levels. Section IV is dedicated to describing Feshbach resonances in isotropic harmonic traps and discusses the applicability of the method to more complicated cases. In Section V we discuss application of our formalism to the problem of two different atoms in a harmonic trap.

\section{Two channel model of a Feshbach resonance}
In our description of Feshbach resonances we follow the two-channel CI model~\cite{Mies2000,Julienne2004}. In this section we briefly review its characteristics in free space. Let us consider two atoms of mass $m_1$ and $m_2$. The Hamiltonian consists of the open collision channel describing a pair of atoms in the spin state $\left|\chi\right>$, the resonant molecular state $\left|n\right>$ in the closed channel and an interchannel coupling, which depends only on the distance between the atoms. By introducing the center of mass and relative motion coordinates
\begin{eqnarray}
\mathbf{R}=\frac{m_1 \mathbf{r}_1+m_2 \mathbf{r}_2}{m_1+m_2}\\
\mathbf{r}=\mathbf{r}_1-\mathbf{r}_2,
\end{eqnarray}
in the absence of external potential the problem can be separated. The center of mass solutions are just the plain waves. The Hamiltonian of the relative motion reads
\begin{equation}
\begin{split}
H=\left|\chi\right>\left<\chi\right|\left(-\frac{\hbar^2}{2\mu}\nabla^2+U_{bg}(\mathbf{r})\right)+\\+\left|n\right>\left<n\right|\left(-\frac{\hbar^2}{2\mu}\nabla^2+U_{mol}(B,\mathbf{r})\right)+\\+\left(\left|\chi\right>\left<n\right|+\left|n\right>\left<\chi\right|\right)W_{n\chi}(r).
\label{ham1}
\end{split}
\end{equation}
Here $\mu$ is the reduced mass, $U$ is the background potential between the atoms away from the resonance and $W_{n\chi}$ is the coupling. The CI wave function is given by
\begin{equation}
\left|\Psi(\epsilon,B,r)\right>=\left|\chi\right>C(\epsilon,B)\Phi_\epsilon (\mathbf{r})+\left|n\right>A(\epsilon,B)\Phi_{mol}(\mathbf{r}),
\end{equation}
where $A$ and $C$ are the amplitudes and $\Phi(r)$ are the channel wave functions, which obey single channel Schr\"{o}dinger equations:
\begin{eqnarray}
\left(-\frac{\hbar^2}{2 \mu}\nabla^2+U_{bg}(\mathbf{r})\right)\Phi_\epsilon (r)=\epsilon \Phi_\epsilon (r),\\
\left(-\frac{\hbar^2}{2 \mu}\nabla^2+U_{mol}(B,r)\right)\Phi_{mol} (r)=\nu(B) \Phi_{mol} (r).
\end{eqnarray}
The resonance is controlled by an external magnetic field $B$. $\nu(B)$ is the energy of the molecule shifted by the presence of the field. The effect of the magnetic field is not so much to change $U_{mol}$, but mostly to control $\nu$. Close to the resonance, $\nu$ may be expanded to first order, giving
\begin{equation}
\nu(B)\approx s(B-B_0),
\end{equation}
where $s$ is the difference of magnetic moments between the open and closed channel states and $B_0$ is the value of the magnetic field at which the energy of the closed channel crosses the dissociation threshold in the open channel. Other important parameters are the resonance width $\Delta$ and background scattering length $a_{bg}$, connected by the identity~\cite{JulienneRMP}
\begin{equation}
s\Delta=\frac{\Gamma(\epsilon)}{2k a_{bg}},
\end{equation}
where $\Gamma$ is the decay width, given by $\Gamma(\epsilon)=2\pi\left|\left<\Phi_{mol}\right|W_{n\chi}\left|\Phi_\epsilon\right>\right|^2$. Within the single-channel description, the scattering length can be obtained, given by the well-known formula~\cite{Timmermans} 
\begin{equation}
a(B)=a_{bg}\left(1-\frac{\Delta}{B-B_{res}}\right),
\end{equation}
This effective expression for the scattering length does not contain the parameter $s$.

\section{Feshbach resonance in a trap}
If the system is in an external trap, the above description needs to be adjusted. First of all, the separation of center of mass and relative motion may no longer be possible. We thus rewrite the full Hamiltonian, adding the trapping potential $U_{trap}$ to the interaction $U$ and $U_{mol}$. The Hamiltonian takes the form
\begin{equation}
\begin{split}
H=\left|\chi\right>\left<\chi\right|\left(T+U_{bg}(r)+U_{trap}(\mathbf{\mathbf{r}_1},\mathbf{r}_2)\right)+\\+\left|n\right>\left<n\right|\left(T+U_{mol}(r)+U_{trap}(\mathbf{r}_1,\mathbf{r}_2)\right)+\\+\left(\left|\chi\right>\left<n\right|+\left|n\right>\left<\chi\right|\right)W_{n\chi}(r),
\end{split}
\end{equation}
where $T=-\frac{\hbar}{2m_1}\nabla_1 ^2-\frac{\hbar}{2m_2}\nabla_2 ^2$ is the kinetic energy operator. The general wave function is given by
\begin{equation}
\left|\Psi(\mathbf{R},\mathbf{r})\right>=\left|\chi\right>\sum_i {C_i \Phi_i (\mathbf{R},\mathbf{r})}+\left|n\right>\sum_k{A_k\Phi_{k\,mol}(\mathbf{R},\mathbf{r})},
\end{equation}
where the channel wave functions obey
\begin{equation}
\left(-\frac{\hbar^2}{2\mu}\nabla^2 _r-\frac{\hbar^2}{2M}\nabla^2 _R +U_{bg}+U_{trap}\right)\Phi_i (\mathbf{r},\mathbf{R})=\epsilon_i \Phi_i (\mathbf{r},\mathbf{R})
\end{equation}
\begin{equation}
\begin{split}
\left(-\frac{\hbar^2}{2\mu}\nabla^2 _r-\frac{\hbar^2}{2M}\nabla^2 _R +U_{mol}+U_{trap}\right)\Phi_{k\,mol} (\mathbf{r},\mathbf{R})=\\=(\nu(B)+\varepsilon_k) \Phi_{k\,mol} (\mathbf{r},\mathbf{R}).
\end{split}
\end{equation}
Previously it was sufficient to consider only one molecular state, but in the presence of the trap one may expect coupling between different free and molecular levels induced by the trapping potential. It is convenient to separate the resonant energy shift from the energy of trap excitations. We will assume that the molecule is a pointlike particle of mass $M=m_1+m_2$, which is a reasonable assumption as long as the interatomic distance is much smaller than characteristic trap lengths. Then the molecular wave function $\Phi_{k\,mol}(\mathbf{R},\mathbf{r})$ can be replaced by its value at $r=0$ and a Dirac delta in $r$. The $R$-dependent part satisfies 
\begin{equation}
\left(-\frac{\hbar^2}{2M}\nabla^2 _z+\tilde{U}_{trap}(\mathbf{R})\right)\Phi_{k\,mol} (\mathbf{R})=(\nu(B)+\varepsilon_k) \Phi_{k\,mol} (\mathbf{R}),
\end{equation}
Here $\tilde{U}_{trap}(\mathbf{R})=U_{trap}(r=0,\mathbf{R})$. Applying the Schr\"{o}dinger equation $H\left|\psi\right>=E\left|\psi\right>$ to this problem gives
\begin{eqnarray}
E\,C_i = \epsilon_i C_i + \sum_l{V^\star _{li} A_l} \label{eq1},\\
E\,A_k = (\nu(B)+\varepsilon_k) A_k+\sum_j{V_{kj} C_j \label{eq2}},
\end{eqnarray}
where $V_{ki}=\left<\Phi_{k\,mol}\right|W_{n\chi}\left|\Phi_i\right>$. The collision takes place at short range in comparison to characteristic trap lengths, which justifies using only one coupling $W_{n\chi}$. By substituting \eqref{eq1} into \eqref{eq2} we get a self-consistent formula for $E$:
\begin{equation}
(E-\nu-\varepsilon_k)A_k=\sum_{jl}{\frac{V_{kj}V^\star _{lj}}{E-\epsilon_j}A_l},
\label{selfc}
\end{equation}
This sum may be divergent, because we treated the molecular state as a pointlike particle with $\delta(r)$ in the relative coordinate. $\nu$ needs then to be renormalized. 

\section{Isotropic harmonic trap}
In this section we apply our formalism to the simplest possible case of two atoms in an isotropic harmonic trap. The trapping potential $\frac{1}{2}m\omega^2 \mathbf{r}_i^2$ separates center of mass and relative motion. We neglect the background interaction in the open channel, assuming that the background scattering length is small~\cite{Marcelis}. The $R$-dependent part of the problem disappears from the equations, meaning that the resonance will not affect the center of mass motion of the pair. Furthermore, due to the form we assumed for the molecular wave function, only states with $\ell=0$ will couple to the closed channel. We thus have
\begin{equation}
\left|\Psi(r)\right>=\left|\chi\right>\sum_j{c_j \phi_j (r)}+\left|n\right>A\Phi_{mol}(r),
\end{equation}
where $\phi_j=\mathcal{N}_j e^{-r^2/2a_{ho}^2}L_j ^{1/2}((r/a_{ho})^2)$, $\mathcal{N}_j=(a_{ho})^{-3/2}\sqrt{\frac{\Gamma(j+1)}{\Gamma(j+3/2)}}$ is the normalization factor and $\Phi_{mol}(r)$ is approximated by $\delta^{(3)}(\mathbf{r})$. Additionally the pair is described by some center of mass wavefunction which does not contribute to the resonance properties.

The coupling between open and closed channel in a harmonic trap may be calculated analytically, using the property $L_j ^{1/2}(0)=\frac{2}{\sqrt{\pi}}\frac{\Gamma(j+3/2)}{\Gamma(j+1)}$. Then 
\begin{equation}
V_j=\int{d^3 r \phi_j(r)W(r)\delta^{(3)}(\mathbf{r})}=\alpha \sqrt{\frac{\Gamma(j+3/2)}{\Gamma(j+1)}},
\end{equation}
where $\alpha$ is a constant. The method to calculate its value in terms of experimentally accessible parameters is given in the Appendix. By inserting this into \eqref{selfc} and denoting $x=(E-3\hbar\omega/2)/2\hbar\omega$, we get
\begin{equation}
E-\nu=-\alpha^2\sum_{n=0}^\infty{\frac{\Gamma(n+3/2)}{\Gamma(n+1)}\frac{1}{n-x}},
\label{notrenormalized}
\end{equation}
where the energy of the molecular state was inserted into $\nu$. The sum in Eq. \eqref{notrenormalized} is divergent. In the numerical calculations, when one uses a finite basis, this results in dependence of the resonance position on the basis size $n^\star$. This can be avoided by renormalizing the parameter $\nu$. The divergence can be extracted by adding and substracting $1/\sqrt{n+1}$ under the sum. It can be shown that 
\begin{equation}
\sum_{n=0}^{n^\star}{\frac{1}{\sqrt{n+1}}}\overset{n^\star\to\infty}{\longrightarrow}\zeta(1/2)+2\sqrt{n^\star},
\end{equation}
where $\zeta$ is the Riemann zeta function. By introducing
\begin{equation}
W(x)=\sum_{n=0}^{\infty}{\left(\frac{\Gamma(n+3/2)}{\Gamma(n+1)}\frac{1}{n-x}-\frac{1}{\sqrt{n+1}}\right)}+\zeta(1/2),
\end{equation}
which is convergent, we obtain
\begin{equation}
E-\nu=-\alpha^2 W(x)-2\alpha^2 \sqrt{n^\star}.
\label{regselfc}
\end{equation}
We may now introduce the renormalized resonance shift $\nu^\star=\nu+2\alpha^2\sqrt{n^\star}$ which makes the equations convergent and ensures that the basis size will not affect the resonance properties. In this form Eq. \eqref{regselfc} is very convenient for numerical calculations. One can also perform the calculations without going to the center of mass frame and using cartesian coordinates, obtaining the same renormalization condition, as done in~\cite{DienerHo2006} in the context of an optical lattice. We note that lower dimensional problems can also be treated with similar approach~\cite{Duan}.

\subsection{Application to nonseparable problems}
The results obtained in the previous paragraphs are not limited to the pure harmonic potential. Instead, they can be used to solve a wider class of problems. Let us now consider a general trapping potential. The eigenfunctions in both channels can be expanded in the basis of harmonic oscillator states
\begin{eqnarray}
\Phi_k(\mathbf{R},\mathbf{r})=\sum_{i}{a^k _{i}\phi_{NLM}(\mathbf{R})\phi_{n\ell m}(\mathbf{r})}\\
\Phi_k ^{mol}(\mathbf{R})=\sum_j{c^k _j \phi_{N'L'M'}(\mathbf{R})},
\end{eqnarray}
where the index $i$ in the sums denotes summation over all possible states $\left|NLMn\ell m\right>$ of the pair and $j$ over molecular states $\left|N'L'M'\right>$. We note that only terms with $L=\ell=0$ will couple to the resonance. By using this basis to perform the calculations, Eq.~\eqref{selfc} takes the form
\begin{equation}
(E-\nu-\varepsilon_k)A_k=\sum_{k'}A_{k'}\sum_{N,N',n,n',t}{a^{t}_{Nn}a^{t\star} _{N'n'} c^{k'}_N c^{k\star}_N\frac{V_{n'}V_n^\star}{E-\epsilon_t}},
\end{equation}
where $\epsilon_t$ ($\varepsilon_k$) denotes the eigenenergies of the open (closed) channel and only expansion coefficients $a$, $c$ with zero angular momentum are present in this formula. As the form of the coupling is the same as in the pure harmonic oscillaotr case, this equation can be renormalized in the same way. This method will be particularily useful when one of the following conditions is met:
\begin{itemize}
\item the trapping potential can be described by harmonic term plus some perturbation with finite strength and range; then the high energy eigenstates will not be affected by the perturbation, or
\item the coupling between center of mass and relative motion mixes only the states lying close to each other.
\end{itemize}
In both cases a reasonably small basis can be used for numerical calculations.
\section{Examplary applications}
\subsection{Two different atoms in harmonic trap}
\begin{figure}
\centering
\includegraphics[width=\columnwidth]{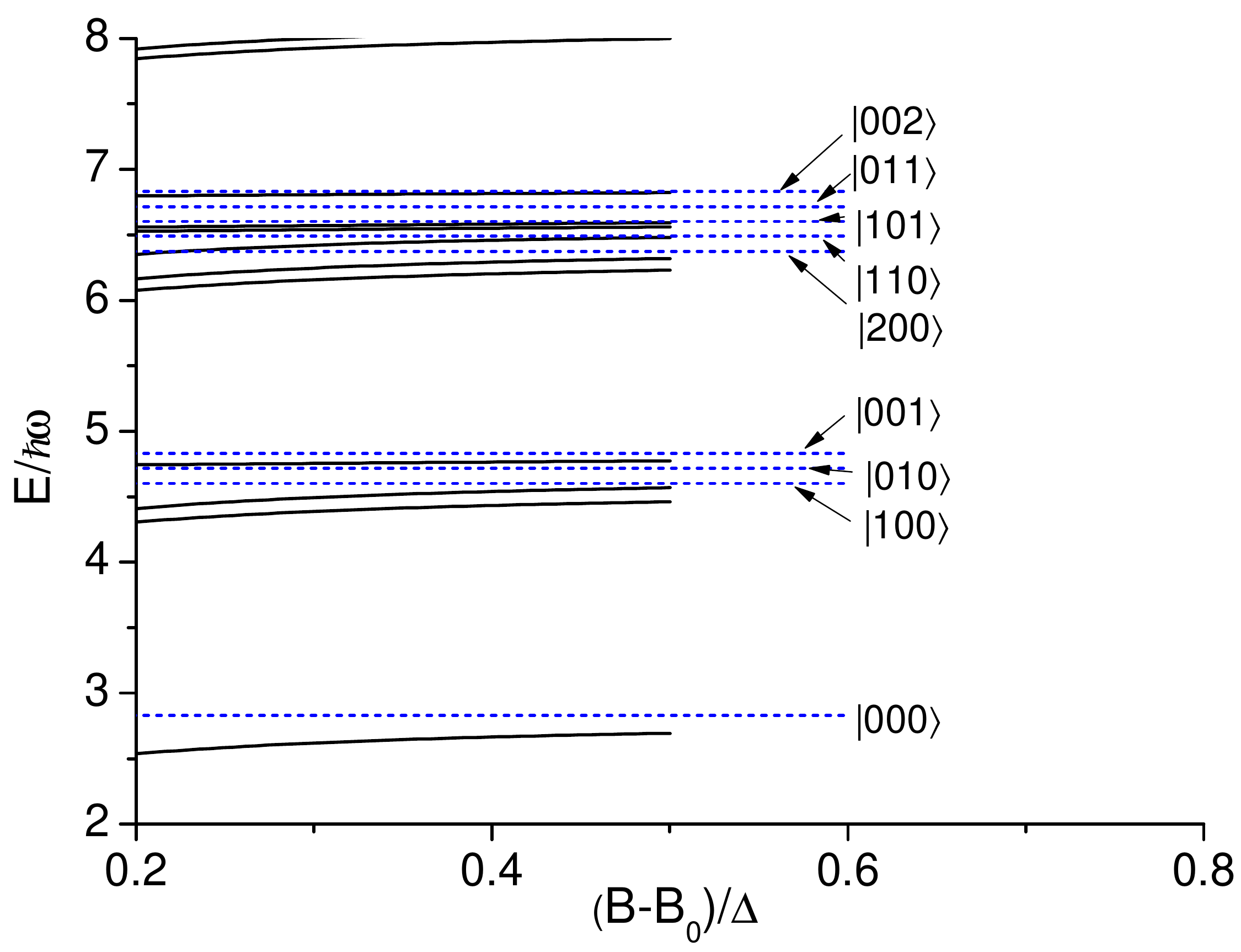}
\caption{\label{enlev}(color online) Energy levels away from the K-Rb resonance. Dashed blue lines depict the uncoupled $C=0$ case, where the eigenstates can be labeled by the quantum numbers $N\ell n$. The actual eigenstates (black solid lines) are composed from them.}
\end{figure}

As a simple example of a system where the center of mass and relative degrees of freedom are coupled, we consider a combination of two different species with masses $m_1$ and $m_2$ in a harmonic trap. Due to different masses and polarizabilities of the atoms, each atom feels different trapping frequencies $\omega_1$ and $\omega_2$. The Hamiltonian of the open channel reads
\begin{equation}
H=-\frac{\hbar^2}{2M}\nabla_R ^2 +\frac{1}{2}M\Omega^2 R^2 -\frac{\hbar^2}{2\mu}\nabla_r ^2 +\frac{1}{2}\mu\omega^2 r^2 +C \mathbf{R}\cdot\mathbf{r},
\end{equation}
where $\Omega=\sqrt{\frac{(m_1 \omega_1 ^2+m_2 \omega_2 ^2)}{M}}$, $\omega=\sqrt{\frac{(m_1 \omega_2 ^2 +m_2 \omega_1 ^2)}{M}}$ and the coupling term $C$ in the Hamiltonian is given by~\cite{Sengstock,Bertelsen2007}
\begin{equation}
C=\mu(\omega_1 ^2-\omega_2 ^2).
\end{equation}
Due to rotational invariance of the Hamiltonian, the total angular momentum $J$ of the pair is conserved and we may choose it to be equal to zero. The open channel wave function may then be expanded in the basis of $J=0$ harmonic oscillator states~\cite{Bertelsen2007}
\begin{equation}
\psi_{N\ell n}(\mathbf{R},\mathbf{r})=\sum_{m=-\ell}^\ell{\frac{(-1)^{\ell-m}}{\sqrt{2\ell+1}}\Phi_{N\ell m}(\mathbf{R})\phi_{n\ell(-m)}(\mathbf{r})},
\label{HRr}
\end{equation}
where $\Phi(\mathbf{R})$ is the center of mass harmonic oscillator wave function, $\phi(\mathbf{r})$ is the relative motion wave function and we used the fact that the Clebsch-Gordan coefficients $\left<\ell_1 M \ell_2 m|00\right>$ give $\frac{(-1)^{\ell_1-m}}{2\ell_1+1}\delta_{\ell_1 \ell_2}$. Matrix elements of the Hamiltonian~\eqref{HRr} in this basis can be computed analytically~\cite{Bertelsen2007} (only the last term in~\eqref{HRr} is not diagonal in this basis). The closed-channel wave function is a superposition of $\ell=0$ eigenstates $\Phi_n$ of a harmonic oscillator with frequency $\Omega$ and mass $M$. We thus have
\begin{equation}
\left|\Psi(\mathbf{R},\mathbf{r})\right>=\left|\chi\right>\sum_k{c_k \psi_k(\mathbf{R},\mathbf{r})}+\left|n\right>\sum_k{A_k \Phi_k(\mathbf{R})\delta(\mathbf{r})},
\end{equation}
where $\psi_k=\sum_{N\ell n}{b^k _{N\ell n}\psi_{N\ell n}(\mathbf{R},\mathbf{r})}$ are the eigenstates of the full Hamiltonian~\eqref{HRr}. Only the $\ell = 0$ components are coupled with the closed channel and the coupling has the same form as in the previous case of single harmonic oscillator, so we can now substitute the wave functions and couplings into~Eq.~(\ref{selfc}) and solve it numerically.
\begin{figure}
\centering
\includegraphics[width=0.5\textwidth]{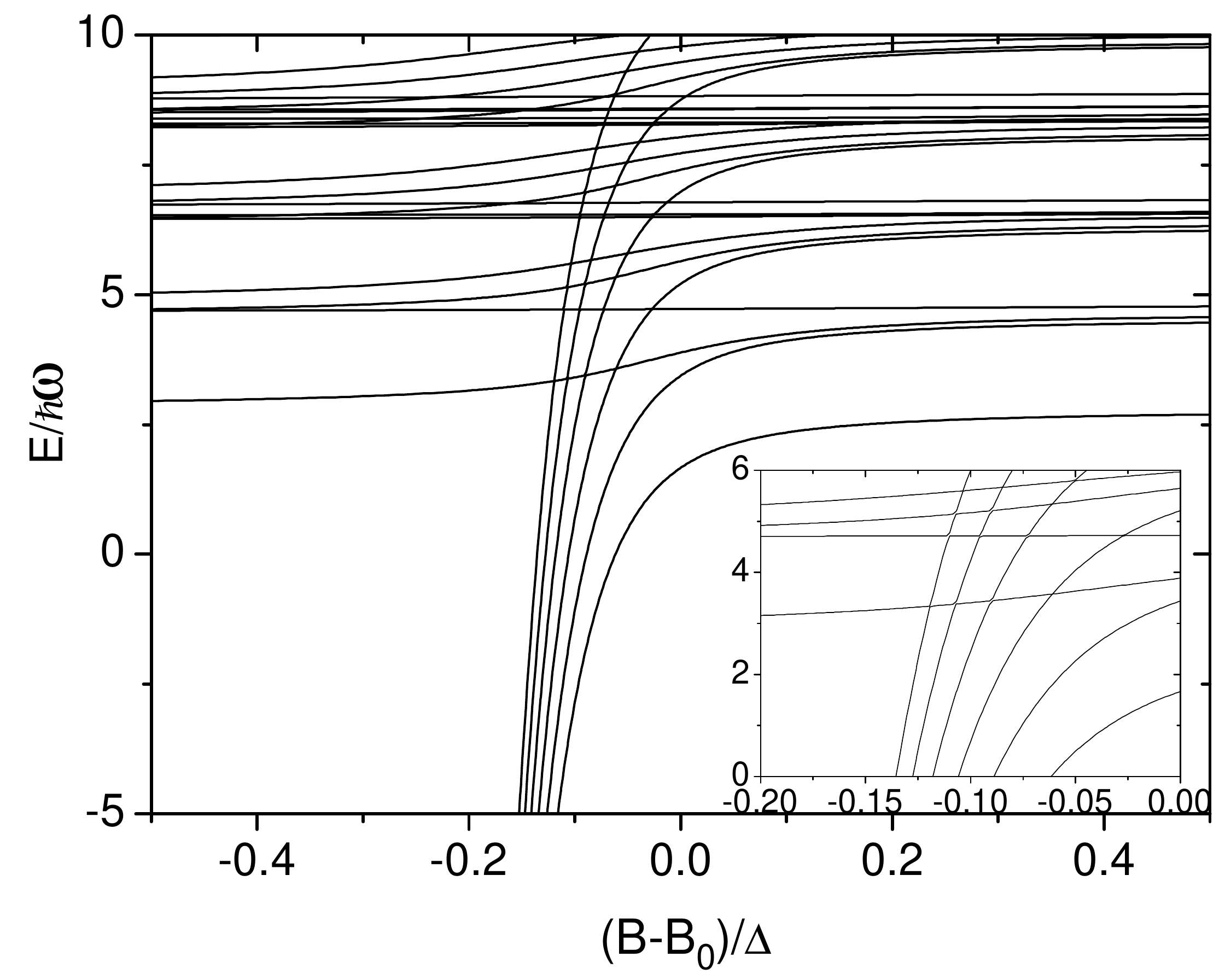}
\caption{\label{KRbenlev}Energy levels in units of the relative trapping frequency $\omega$ for a K-Rb Feshbach resonance. The ratio of the trapping frequencies between K and Rb atoms is assumed to be $1.4$ as in \cite{Sengstock}. Five molecular bound states are taken into account. The inset shows a closer view of the region where the bound states cross the trap levels.}
\end{figure}
We now analyze some particular examples of heteronuclear Feshbach resonances. Fig. \ref{KRbenlev} shows the energy levels in the case of a Feshbach resonance between K and Rb at $547$~G~\cite{Sengstock}, assuming the trapping frequency for rubidium $\omega_1=10$~kHz and for potassium $\omega_2=14$~kHz. The parameters of the resonance can be found in~\cite{JulienneRMP}. Away from resonance the eigenstates do not contain any bound levels. The energies for this case are shown on Fig. \ref{enlev}. We note that due to coupling of motional degrees of freedom induced by nonzero $C$, the eigenstates are composed out of several harmonic oscillator levels with different angular momenta and the corrections to the eigenenergies with respect to the uncoupled case are significant. Close to the resonance one can see the deeply bound states to which one can assign the quantum number $N$ labelling the trap level. Then the molecular bound states cross with the free atomic states, as shown by the inset of Fig.~\ref{KRbenlev}. Due to the different symmetry of the states, we can expect that these are true level crossings. To verify this, we checked numerically that the crossing states are orthogonal.

\begin{figure}
\centering
\includegraphics[width=0.5\textwidth]{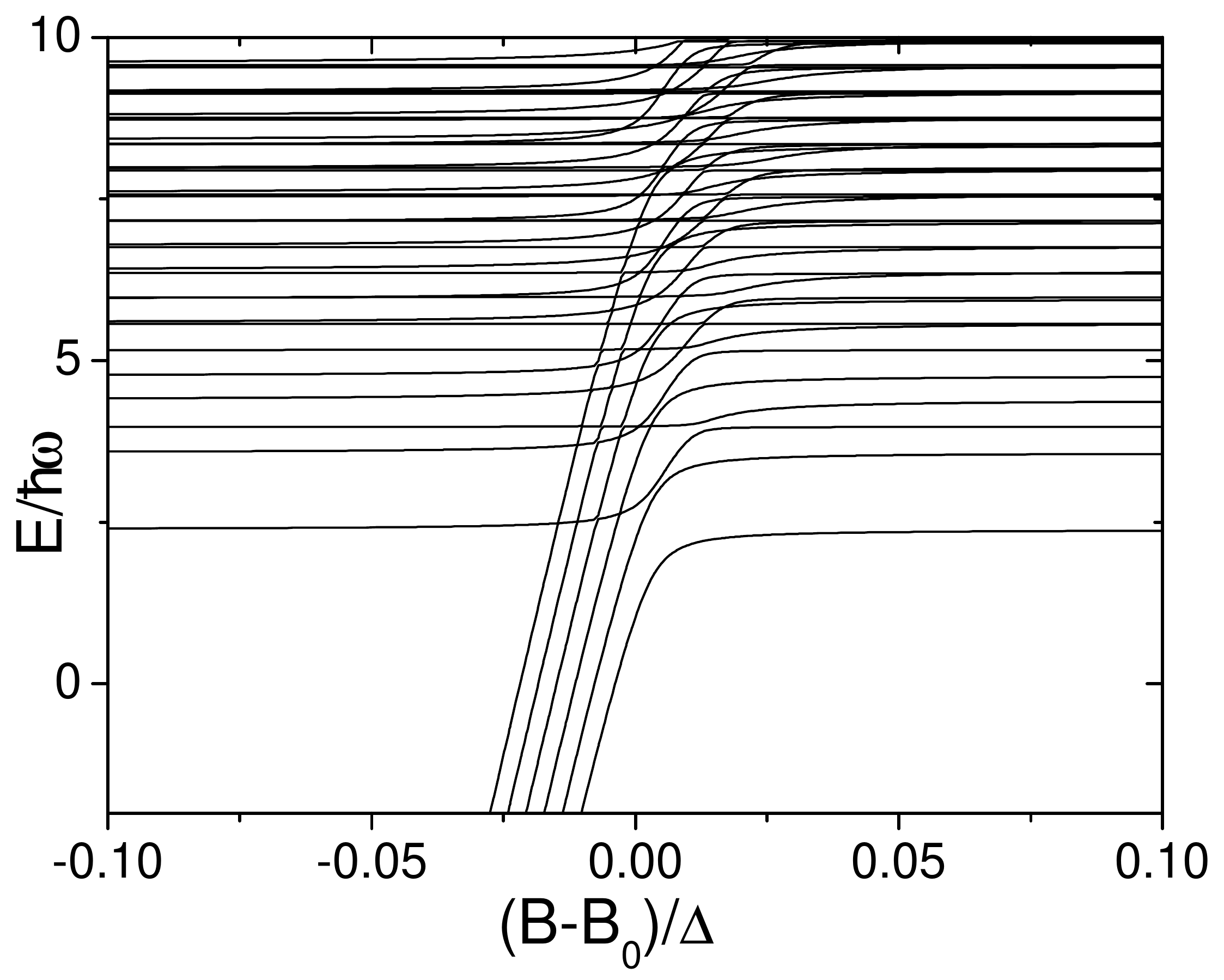}
\caption{\label{LiCsenlev}Same as on Fig.~\ref{KRbenlev}, but for Li-Cs resonance, where the ratio of trapping frequencies is assumed to be $1.8$.}
\end{figure}

\begin{figure}
\centering
\includegraphics[width=0.5\textwidth]{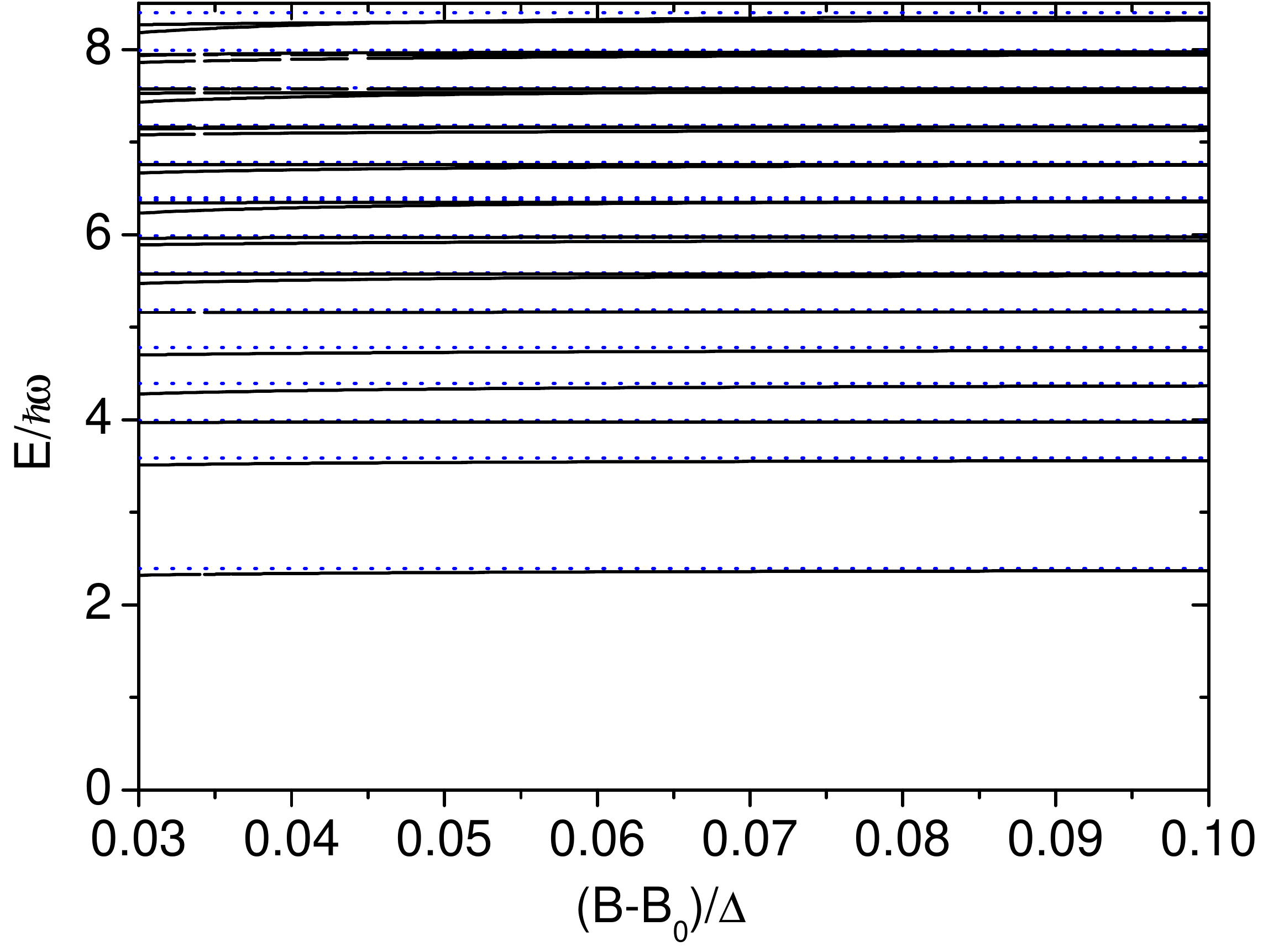}
\caption{\label{LiCsaway}(color online) Same as on Fig.~\ref{enlev}, but for Li-Cs resonance. The coupling term plays less important role than in the K-Rb case.}
\end{figure}

In Fig.~\ref{LiCsenlev} we present the case of Li-Cs resonance at $B=816$~G, which has recently been observed experimentally~\cite{Weidemueller2012}. Here we assumed that the trapping frequency is $\omega_1=1$~kHz for Cs atoms and $\omega_2=1.8$~kHz for lithium. Due to the large mass difference, the ratio of trapping frequencies here is bigger than in the K-Rb case. As a resul, in the former case the energy levels away from the resonance tended to form groups, but here it is not the case. Instead we get an energy spectrum which looks more complicated, but has similar nature as before. The coupling of center of mass and relative motion occurs to have less impact than in the K-Rb case, so away from resonance the eigenstates are less distorted from the pure $\psi_{N\ell n}$ states (see Fig.~\ref{LiCsaway}).

\section{Conclusion}
In this paper, we presented a general formalism describing Feshbach resonances in an external trap. Our method works for both open- and closed channel dominated resonances and can be applied to nonseparable traps. It is particularly efficient when the single particle trap eigenfunctions can be expanded in harmonic oscillator basis. We presented results for the calculation of energy levels of a pair of different atoms, where the trapping frequencies cannot be assumed to be the same. Apart from static cases, our formalism allows for calculation of the dynamics of the wave functions where the trap parameters or magnetic field are changing in time. This can be useful for example for quantum computations, where control of the qubits will be enhanced by Feshbach resonances.

This work was supported by the Foundation for Polish Science International PhD Projects Programme co-financed by the EU European Regional Development Fund, AQUTE, SFB/TRR21 and National Center for Science Grant No. DEC-2011/01/B/ST2/02030.
\appendix
\section{Relation between coupling constant $\alpha$ and resonance parameters}
The connection between $\alpha$ and experimentally accessible parameters can be found by comparison of Eq.~\ref{regselfc} with the energy of a weakly bound molecule in free space, where $E=-\frac{\hbar^2}{2 \mu a^2}$ and $a(E)$ is the effective energy dependent scattering length \cite{JulienneRMP}. This can be done by taking the limit $\omega\rightarrow 0$ at constant $E$. In this limit $x\rightarrow -\infty$ and $W(x)\rightarrow -\pi\sqrt{-x}$. Equation \eqref{regselfc} takes the form
\begin{equation}
E-\nu^\star=\pi \tilde{\alpha}^2\left(\frac{\mu}{2\hbar^2}\right)^{3/2}\sqrt{-E}.
\end{equation}
here $\alpha^2$ was rescaled as $\tilde{\alpha}^2 a_{ho}^{-3}$, where $a_{ho}=\sqrt{\hbar/\mu\omega}$. Expanding $a(E)$ into power series according to the effective range theory~\cite{JulienneRMP}
\begin{equation}
\frac{1}{a(E)}= \frac{1}{a_{bg}}-\frac{1}{2} r_0 \frac{2\mu E}{\hbar^2}+\ldots,
\label{effrange}
\end{equation}
where $r_0$ is the effective range parameter and $a_{bg}$ is the s-wave scattering length away from the resonance, we obtain the equation
\begin{equation}
(E-\nu^\star)\left(\frac{2\hbar^2}{\mu}\right)\frac{1}{\pi \tilde{\alpha}^2}=\frac{\hbar}{\sqrt{2\mu}}\frac{1}{a_{bg}}-\frac{\mu}{\sqrt{2}\hbar}r_0 E.
\end{equation}
Comparing the energy-dependent and independent terms, we conclude that
\begin{eqnarray}
\nu^\star=\frac{\hbar^2}{\mu a_{bg} r_0}
\label{nu}\\
\tilde{\alpha}=\frac{2\hbar^2}{\mu\sqrt{\pi}}\sqrt{-\frac{1}{r_0}}.
\label{alpha}
\end{eqnarray}
We notice that the effective range for this problem is negative. It can be found using the definition of energy-dependent scattering length in the presence of Feshbach resonance \cite{Bolda2002,Idziaszek2006}
\begin{equation}
a(E)=a_{bg}\left(1-\frac{\Delta(1+E/E_b)}{B-B_0+\Delta E/E_b-E/s}\right),
\label{scattlength}
\end{equation}
where $E_b$ is the bound state energy. At $E\approx0$ this reduces to the common formula $a=a_{bg}\left(1-\frac{\Delta}{B-B_0}\right)$. Expanding \eqref{scattlength} in $E$, we get
\begin{equation}
\frac{1}{a(E)}=\frac{1}{a_{bg}}+\frac{E_b \Delta+(B-B_0)s\Delta-s\Delta^2}{a_{bg}E_b s (B-B_0-\Delta)^2}E+\ldots
\end{equation}
Comparing this with \eqref{effrange} and assuming that $B$ is close to $B_0$, we get
\begin{equation}
r_0=\frac{\hbar^2}{\mu a_{bg} s(B-B_0)}.
\end{equation}
Inserting this result into \eqref{nu} and \eqref{alpha} and neglecting the contribution from background scattering length (as we are close to resonance), we get
\begin{eqnarray}
\nu^\star=s(B-B_0)\\
\tilde{\alpha}=2\hbar\sqrt{\frac{a_{bg}s\Delta}{\mu\pi}}.
\end{eqnarray}
\bibliography{allarticles}
\end{document}